%% file: main.tex
\documentclass[copyright,submission]{eptcs}
\providecommand{\event}{FMAS 2024}

\usepackage[T1]{fontenc}
\usepackage[utf8]{inputenc}

\usepackage{amsmath}
\usepackage{amssymb}

\usepackage{bm}
\renewcommand{\vec}{\boldsymbol}

\usepackage{siunitx}

\usepackage[cal=euler]{mathalfa} 

\usepackage{graphicx}
\usepackage{caption}
\usepackage{subcaption}

\usepackage{listings}
\usepackage{tikz}
\tikzset{
  every picture/.style={%
    every node/.style={font=\small},
    every path/.style={%
      shorten >=1pt,
      shorten <=1pt,
    }
  }
}

\input{tikz_settings}

\usepackage{microtype}

\usepackage{todonotes}

\usepackage[capitalise]{cleveref}

\usepackage{soul}

\usepackage[inline]{enumitem}
\usepackage{xcolor}
\usepackage{multirow}

\title{Creating a Formally Verified Neural Network\\for Autonomous Navigation: An Experience Report\thanks{This publication has emanated from research conducted with the financial support of Science Foundation Ireland under Grant number 20/FFP-P/8853.}}
\author{
  Syed Ali Asadullah Bukhari
  \qquad
  Thomas Flinkow
  \qquad
  Medet Inkarbekov
  \\
  Barak A. Pearlmutter
  \qquad
  Rosemary Monahan
  \institute{Department of Computer Science, Maynooth University, Maynooth, Ireland}
  \email{\{ali.bukhari,thomas.flinkow,medet.inkarbekov,rosemary.monahan\}@mu.ie\quad barak@pearlmutter.net}
}

\def\titlerunning{Creating a Formally Verified Neural Network for Autonomous Navigation: An Experience Report}
\def\authorrunning{S. A. A. Bukhari, T. Flinkow, M. Inkarbekov, B. A. Pearlmutter \& R. Monahan}

\hypersetup{
  bookmarksnumbered,
  pdftitle    = {\titlerunning},
  pdfauthor   = {\authorrunning},
}

\begin{document}
\maketitle

\begin{abstract}
The increased reliance of self-driving vehicles on neural networks opens up the challenge of their verification. 
In this paper we present an experience report, describing a case study which we undertook to explore the design and training of a neural network on a custom dataset for vision-based autonomous navigation.
We are particularly interested in the use of machine learning with differentiable logics to obtain networks satisfying basic safety properties by design, guaranteeing the behaviour of the neural network after training.
We motivate the choice of a suitable neural network verifier for our purposes and report our observations on the use of neural network verifiers for self-driving systems.
\end{abstract}

\section{Introduction and Motivation}
A shift from a purely rule-based software to a learning-based approach for control of autonomous driving systems is evident in recent years \cite{Pomerleau-1988a, ma2020artificial}.
This change can be attributed primarily to the advances in Deep Neural Networks (DNNs) and their improved ability to handle the complexity of environments compared to conventional autonomous navigation methods. 
Moreover, the availability of hardware accelerators and GPUs on edge devices at low cost and power has also progressed the use of DNNs for these systems.
In addition, the computing ability coupled with the presence of various on-device sensors, such as Lidar and cameras, makes it possible to achieve the task of controlling vehicles without the need of human intervention \cite{Thrun-2010a}.
In particular, vision-based systems have been successful for this purpose as on-board cameras can be used to help auto navigation as well as providing real time video monitoring, as is often required in these systems.
Some notable examples of vision-based systems employing DNNs include obstacle avoidance, path following, and object detection \cite{ravindran2020multi, ZHAO2024122836}.

The increased reliance of self-driving vehicles on neural networks opens up the challenge of their verification.
The safety-critical nature of these vehicles calls for their complete verification before they are put to use in a real environment.
In general, the verification of any safety-critical component in any system is viewed as a crucial task, as an overlooked corner case may lead to a catastrophic condition or an irreparable loss.
However, the presence of neural networks in a system makes the verification task further challenging \cite{seshia2022toward}. 

Many efforts have been made to verify neural networks, with verifiers like VNN~\cite{tranNNVNeuralNetwork2020}, Marabou~\cite{katzMarabouFrameworkVerification2019} and $\alpha,\beta$-CROWN~\cite{zhang2018efficient,xu2020automatic,xu2021fast,wang2021beta,zhang2022general,kotha2023provably,shi2024genbab} (the winner of the recent VNN-COMP neural network verification competitions \cite{bakSecondInternationalVerification2021,mullerThirdInternationalVerification2022,brixFourthInternationalVerification2023}).
These verifiers have shown promising results for verification of robustness properties of neural network models used for objects classification.
Robustness, in the case of image classification, can be described as the ability of the neural network to retain the prediction label of an input image in response to a small change in the input image.

While the aforementioned verifiers usually assume a trained network with fixed weights, another area of research is the design of correct-by-construction neural networks.
One approach in this direction is \emph{differentiable logics}~\cite{fischerDL2TrainingQuerying2019,vankriekenAnalyzingDifferentiableFuzzy2022,slusarzLogicDifferentiableLogics2023a}. The fundamental principle of machine learning is to minimise a so-called \emph{loss function} which indicates how wrong the network output is compared to the desired output. Differentiable logics are used to  transform a logical constraint $\phi$ into an additional logical loss term $\mathcal{L}_\phi$ to minimise when learning. This loss term is in addition to the standard loss (such as mean-squared error loss $\mathcal{L}_\text{MSE}$). Therefore, the optimisation objective is of the form $\mathcal{L}=\mathcal{L}_\text{MSE}+\lambda\mathcal{L}_\phi$.
It is important to balance the different loss terms, and for this we employ gradient normalisation (GradNorm)~\cite{chenGradNormGradientNormalization2018}, an adaptive loss balancing approach which treats $\lambda$ as an additional learnable parameter $\lambda(t)$. GradNorm has been shown to outperform grid search, the conventional algorithm used in machine learning for hyperparameter tuning.

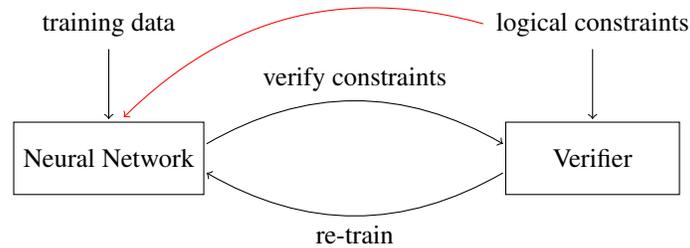
\begin{figure}
    \centering
    \begin{tikzpicture}    
      \tikzstyle{block} = [draw, rectangle, minimum height=2.5em, minimum width=6em]

      \node [block] (network) {Neural Network};
      \node [block, right=4 cm of network] (verifier) {Verifier};

      \node [above=1cm of network] (data) {training data};
      \node [above=1cm of verifier] (constraints) {logical constraints};

      \draw[->, bend left] ([yshift=5pt]network.east) to node[midway, above, align=center] {verify constraints} ([yshift=5pt]verifier.west);
      \draw[->, bend left] ([yshift=-5pt]verifier.west) to node[midway, below, align=center] {re-train} ([yshift=-5pt]network.east);

      \draw[->] (data) -- (network);
      \draw[->] (constraints) -- (verifier);

      \draw[->, red, bend right] (constraints.west) to ([xshift=5pt, yshift=1pt]network.north);
    \end{tikzpicture}
    \caption{Differentiable logics allow for the use of logical constraints during training by translating them into additional loss terms. Note that in this paper, we do not use continuous verification, that is, we do not re-train the network after the verification; instead, we only try to evaluate what influence training with logical constraints has on the verification afterwards.}
    \label{fig:differentiable_logics}
\end{figure}

\Cref{fig:differentiable_logics} shows the standard loop of training a network, verifying it, and re-training the network if necessary.
Differentiable logics allow the integration of constraints as additional loss terms into the training process. Multiple mappings have been defined in the literature to translate logical constraints into loss terms, which allow  for real-valued truth values, but are also differentiable almost everywhere for use with standard gradient-based methods.
Prominent examples include DL2~\cite{fischerDL2TrainingQuerying2019} and fuzzy logic based mappings~\cite{vankriekenAnalyzingDifferentiableFuzzy2022, slusarzLogicDifferentiableLogics2023a}. 
In our experiments, we use the Gödel fuzzy logic $[\![\cdot]\!]:[0,1]\to[0,1]$ which translates as follows:
\begin{itemize}
   \item  conjunction as $[\![x\land y]\!]=\min(x,y)$, 
      \item  disjunction as $[\![x\lor y]\!]=\max(x,y)$, and
      \item  implication as $[\![x\rightarrow y]\!]=\begin{cases}1,&x<y\\y,&\text{else}.\end{cases}$
\end{itemize}
We chose to use this logic translation as it provides a desirable property for the translation process i.e. that the operators have strong derivatives almost everywhere.
Previous experimental results~\cite{flinkowComparingDifferentiableLogics2024} suggest that the choice of a logic does not have a major impact.

In this paper we present an experience report, describing a case study which we undertook to explore the following challenges:
\begin{enumerate*}[label=(\arabic*)]
    \item design and training of a neural network on a custom dataset for vision-based autonomous navigation,
    \item use of machine learning with differentiable logics to obtain networks satisfying basic safety properties,
    \item obtaining formal guarantees with formal verification of the neural network after training.
\end{enumerate*}
We motivate the choice of a suitable neural network verifier for our purposes and report our observations on the use of neural network verifiers for self-driving systems. As can be seen in~\cref{sec:related_work}, to the best of our knowledge, such a setup of using formal methods at multiple stages of the machine learning pipeline has not been published before.

\section{Related Work}\label{sec:related_work}
The verification of autonomous driving systems, especially on edge devices, remains a largely under-explored area. Despite the rapid progress in autonomous vehicle technology \cite{padmaja2023exploration}, few studies have addressed the verification challenges in resource-constrained environments. This section provides an overview of some foundational contributions in this domain.

Sun et al.~\cite{sun2019formal} present a framework for formal verification of safety properties in autonomous systems controlled by neural networks.
Their approach focuses on ensuring that a robot can safely navigate environments with polyhedral obstacles by constructing a finite state abstraction of the system and applying reachability analysis.
The introduction of imaging-adapted partitions in their work is particularly relevant to the verification of robotic cars, as it simplifies the modelling of LiDAR-based perception, making the verification process more manageable.

Habeeb et al.~\cite{habeeb2023verification} develop a procedure for the safety verification of camera-based autonomous systems, which includes a falsification approach that collects unsafe trajectories to assist in retraining neural network controllers.
Their concept of image-invariant regions is noteworthy, as it enables reasoning about trajectories at the level of regions rather than individual positions, potentially reducing the complexity of the verification process for systems relying on visual inputs.

Cleaveland et al.~\cite{cleaveland2022risk} propose a risk verification framework for stochastic systems with neural network controllers, focusing on estimating the risk of failure under stochastic conditions.
Their work is significant in handling the verification of systems where environmental perturbations might occur, which is critical for autonomous vehicles operating in unpredictable real-world environments.
The empirical validation of their framework using autonomous vehicles demonstrates its applicability to scenarios where robustness to environmental changes is essential.

Ivanov et al.~\cite{ivanov2020case} present a benchmark for assessing the scalability of verification tools in the context of an autonomous racing car controlled by a neural network.
Their work highlights the challenges in verifying systems with high-dimensional inputs, such as LiDAR data, and underscores the importance of addressing the gap between simulated and real-world performance, known as the sim2real gap, in verification efforts.
The limitations identified in their study, particularly regarding sensor faults in real-world environments, emphasise the need for more scalable and robust verification methods.

While these approaches address key verification aspects, they fall short in addressing the specific requirements for edge devices used in autonomous driving systems. Further research is necessary to bridge this gap and provide effective solutions for resource-constrained environments.

\section{Case Study: A Verified Neural Network for Autonomous Navigation}
The goal of this case study is to build a formally verified regression neural network that detects the centre of the track on which the vehicle is travelling.
We train the network in two different ways. First, we  train in a standard manner, training only on the dataset. Our second approach aims to achieve a correct-by-construction network by training using the constraints that we want the network to satisfy after training. In this approach we use differentiable logics.

\subsection{Experimental Setup}
For demonstration purposes, we have used the JetBot kit available at~\cite{sparkfun}. JetBot~\cite{jetbot} is an open-source AI robotic vehicle based on NVIDIA Jetson Nano~\cite{jetsonnano}. Jetson Nano is an edge device equipped with GPUs to train and run neural networks at a lower power range of \qtyrange{5}{10}{\watt}.
It can be easily interfaced with several sensor modules including a camera, making it suitable for demonstrating a vision-based autonomous navigation application.
The JetBot kit features a Leopard Imaging 136 FOV Camera~\cite{jetbotcamera} ($3280{\times}2464$ pixels).

\subsection{Data Collection}
Various neural network models and training data are readily available for a range of computer vision tasks, thus providing a possible foundation for development of vision-based autonomous navigation systems. Additionally, there are multiple open source architectures for autonomous car navigation such as Openpilot~\cite{CommaaiOpenpilot2024}.  A lane-keeping dataset for autonomous plane taxiing was presented in~\cite{fremontFormalAnalysisRedesign2020a}, where they also verified the neural network component. In this case study, we generate our own dataset as our focus is on a meaningful prototype, rather than developing a state-of-the-art lane keeping autonomous vehicle. We build upon the example provided with JetBot, available at~\cite{jetbotroadfollowing}.

\begin{figure}
    \centering
    \begin{subfigure}{0.3\textwidth}
        \centering
        \includegraphics[width=0.9\linewidth, trim=70 20 90 38, clip]{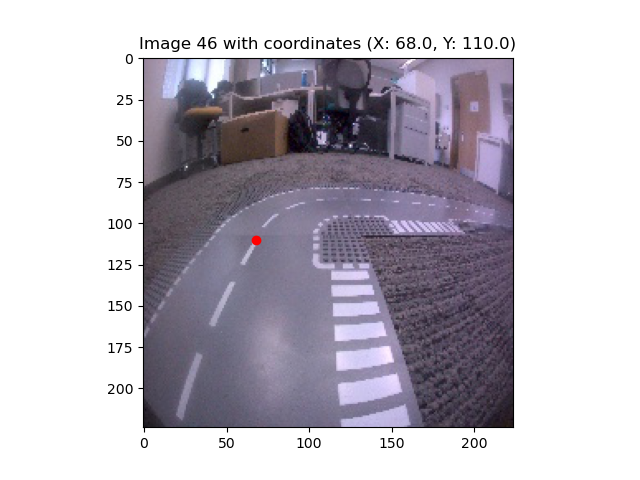}
        \caption{label = 68 ,110}
    \end{subfigure}
    \hfill
    \begin{subfigure}{0.3\textwidth}
        \centering
    \includegraphics[width=0.9\linewidth, trim=70 20 90 38, clip]{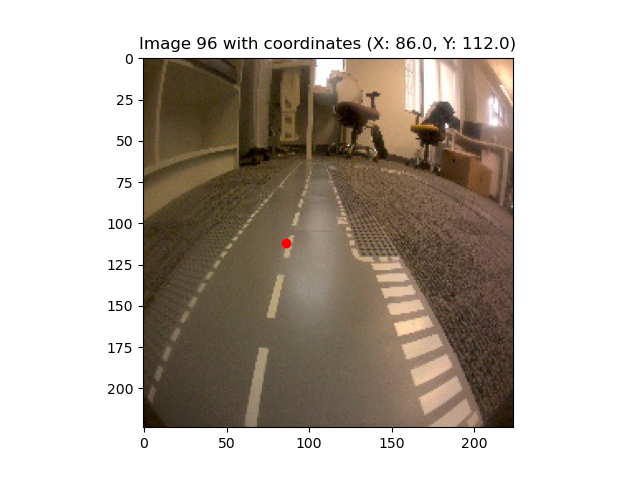}
        \caption{label = 86, 112}    
    \end{subfigure}
    \hfill
    \begin{subfigure}{0.3\textwidth}
        \centering
        \includegraphics[width=0.9\linewidth, trim=70 20 90 38, clip]{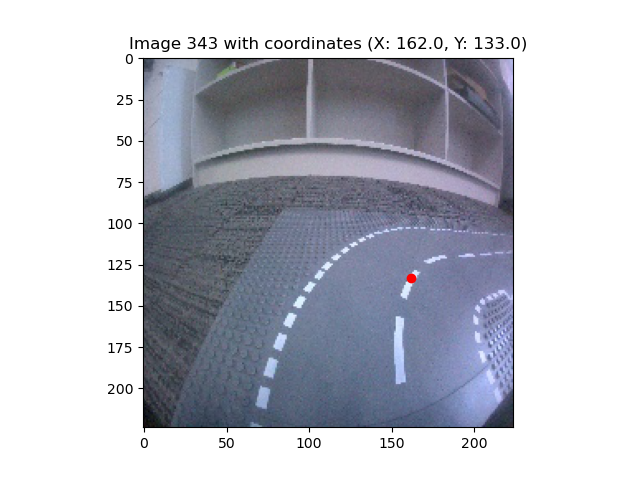}
        \caption{label = 162, 133}
    \end{subfigure}
    \caption{Samples from the LEGO road dataset showing various lighting conditions and track configurations. The red dots located at label coordinates show the centre of the path to be followed in the frame.}
    \label{fig:lego}
\end{figure}

More specifically, we have built a reference track using readily available Lego road plates (with a centred line).
The data is collected by placing the robotic car at various positions on the tracks and taking images with the onboard camera facing down on the track. The images are labelled with $x$ and $y$ coordinates indicating the centre of the path to be followed in the frame. These coordinates are used to calculate the driving parameters for the JetBot motor.
The dataset consists of 385 images stored in a resolution of $224{\times}224$ pixels.
A few samples from the dataset are shown in~\cref{fig:lego}.
The dataset is publicly available at \url{https://github.com/tflinkow/fmas2024}

\subsection{Neural Network Architecture}
Scalability is an important issue for neural network verifiers due to the high-dimensional inputs, the large number of neurons in the network, and the use of non-linear activation functions. Hence, in order to make verification tractable, finding a sufficiently small neural network architecture is essential in this case study. For that reason, we first chose to downscale the original $224{\times}224$ pixel RGB images to $112{\times}112$ pixels and convert them to grey scale before passing them to the network, thus allowing for a noticeably smaller network architecture.

\begin{figure}
  \centering
  \includegraphics[width=\linewidth]{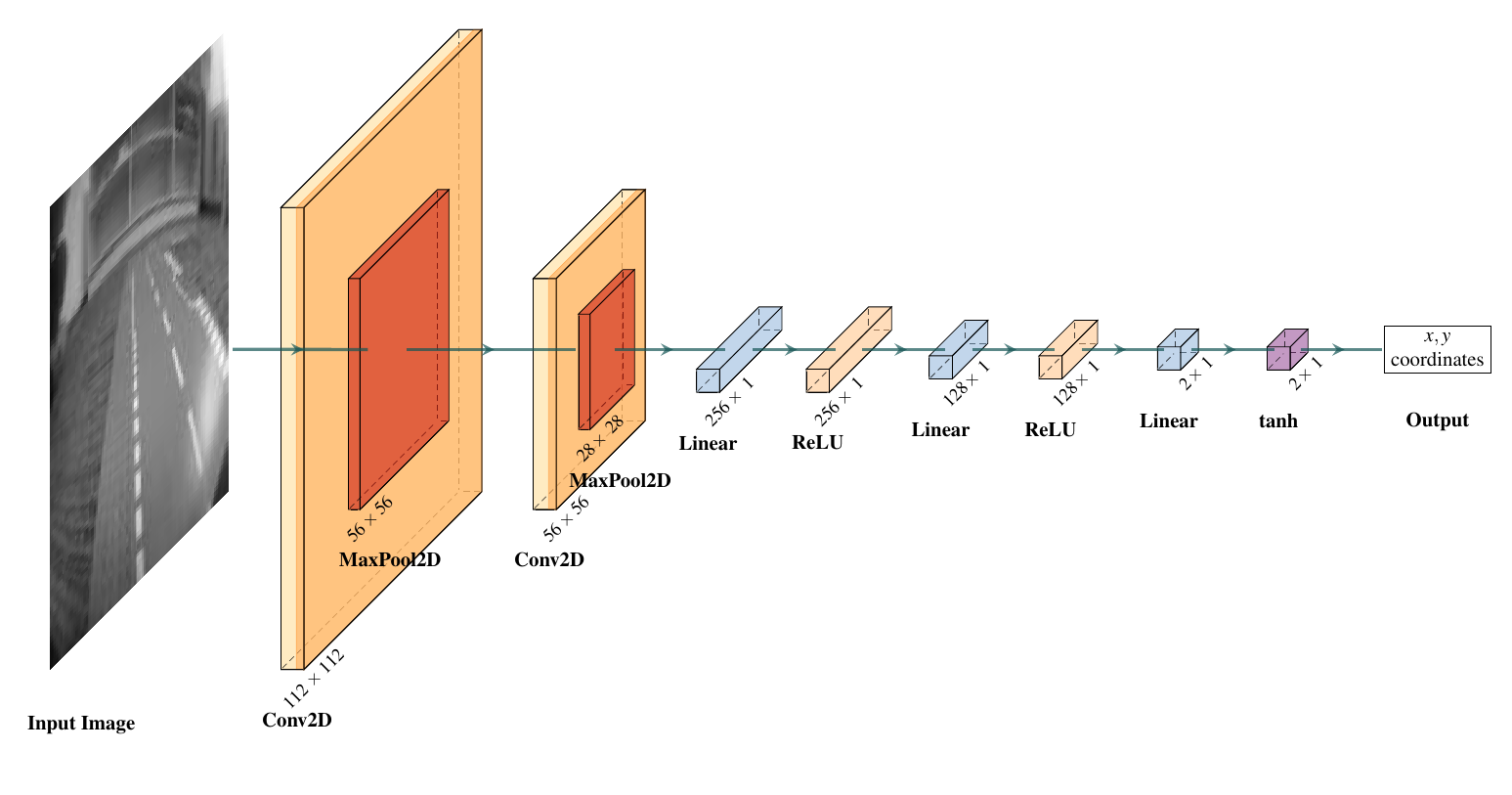}
    \caption{The proposed neural network architecture for extracting the path centre in an input image of size $112\times 112$ pixels.}
    \label{fig:NN}
\end{figure}

The neural network we propose consists of two convolutional layers with ReLU activations (each followed by a max-pooling layer) followed by three fully connected layers, the first two with ReLU activations, and the last layer followed by a $\tanh$ activation.
Further details about each layer and other network parameters are listed in Table \ref{table:t2}.
The network contains 234,136 parameters.
Our proposed architecture is shown in~\cref{fig:NN}.

In the following, we consider our network to approximate the function
\begin{equation}
    \mathcal{N}:\mathbb{R}^{112\times 112}\to\mathbb{\mathbb{R}}^2,   
\end{equation}
where the inputs are images of size $112\times 112$, and the outputs are the $x,y$-coordinates of the centre of the track in the image (according to the image label).

\begin{table}
\caption{Summary of the proposed neural network.}
\begin{tabular}{|l|cc|c|c|}
\hline
\multicolumn{1}{|c|}{\multirow{2}{*}{\textbf{\begin{tabular}[c]{@{}c@{}}Network\\ Layer\end{tabular}}}} &
  \multicolumn{1}{c|}{\textbf{Input Shape}} &
  \multicolumn{1}{c|}{\textbf{Output Shape}} &
  \textbf{Kernel (if applicable)} &
  \multicolumn{1}{c|}{\multirow{2}{*}{\textbf{\begin{tabular}[c]{@{}c@{}}No.\ of Trainable \\ Parameters\end{tabular}}}} \\ \cline{2-4}
\multicolumn{1}{|c|}{} &
  \multicolumn{2}{c|}{\textbf{$\textit{Size} \times \textit{Size} \times \textit{Channel} \times \textit{Batch}$}} &
  \textbf{Size, Stride, Padding} &
  \multicolumn{1}{c|}{} \\ \hline
Conv2D &
  \multicolumn{1}{c|}{$112 \times 112 \times 1 \times 1$} &
  $112 \times 112 \times 1 \times 1$ &
  \multicolumn{1}{c|}{$3 \times 3$,  $1 \times 1$, $1 \times 1$} &
  10 \\ \hline
ReLU &
  \multicolumn{1}{c|}{$112 \times 112 \times 1 \times 1$} &
  $112 \times 112 \times 1 \times 1$ &
  --- &
  --- \\ \hline
MaxPool2D &
  \multicolumn{1}{c|}{$112 \times 112 \times 1 \times 1$} &
  $56 \times 56 \times 1 \times 1$ &
  \multicolumn{1}{c|}{$2 \times 2$,  $2 \times 2$, none} &
  --- \\ \hline
Conv2D &
  \multicolumn{1}{c|}{$56 \times 56 \times 1 \times 1$} &
  $56 \times 56 \times 1 \times 1$ &
  \multicolumn{1}{c|}{$3 \times 3$,  $1 \times 1$, $1 \times 1$} &
  10 \\ \hline
ReLU &
  \multicolumn{1}{c|}{$56 \times 56 \times 1 \times 1$} &
  $56 \times 56 \times 1 \times 1$ &
  --- &
  --- \\ \hline
MaxPool2D &
  \multicolumn{1}{c|}{$56 \times 56 \times 1 \times 1$} &
  $28 \times 28 \times 1 \times 1$ &
  \multicolumn{1}{c|}{$2 \times 2$,  $2 \times 2$, none} &
  --- \\ \hline
Linear &
  \multicolumn{1}{c|}{$784 \times 1$} &
  $256 \times 1$ &
  --- &
  200, 960 \\ \hline
ReLU &
  \multicolumn{1}{c|}{$256 \times 1$} &
  $256 \times 1$ &
  --- &
  --- \\ \hline
Linear &
  \multicolumn{1}{c|}{$256 \times 1$} &
  $128 \times 1$ &
  --- &
  32, 896 \\ \hline
ReLU &
  \multicolumn{1}{c|}{$128 \times 1$} &
  $128 \times 1$ &
  --- &
  --- \\ \hline
Linear &
  \multicolumn{1}{c|}{$128 \times 1$} &
  $2  \times 1$ &
  --- &
  258 \\ \hline
Tanh &
  \multicolumn{1}{c|}{$2  \times 1$} &
  $2  \times 1$ &
  --- &
  --- \\ \hline
\end{tabular}
\label{table:t2}
\end{table}

\subsection{Verification Properties}
The property we wish to verify is a standard local robustness property~\cite{casadioNeuralNetworkRobustness2022} for neural networks, formally defined as

\begin{equation}\label{eq:robustness}
    \mathsf{Robustness}(\vec{x}_0, \epsilon, \delta):=\forall \vec{x}\ldotp ||\vec{x}_0-\vec{x}||_{\infty}\leq\epsilon \implies ||\mathcal{N}(\vec{x}_0)-\mathcal{N}(\vec{x})||_{\infty}\leq \delta.
\end{equation}

This property checks that for slight perturbations $\vec{x}$ within some bound $\epsilon$ of a given input image $\vec{x}_0$, the neural network $\mathcal{N}$ should give roughly the same output, i.e., the measure of difference between the network's outputs $\mathcal{N}(\vec{x}_0)$ and $\mathcal{N}(\vec{x})$ should be within an acceptable threshold $\delta$.

As explained in~\cite{fischerDL2TrainingQuerying2019}, learning properties of the form $\forall\vec{x}\ldotp||\vec{x}_0-\vec{x}||_{\infty}\le\epsilon\implies\phi$ can be approximated by finding a counterexample (i.e., the worst perturbation) $\vec{x}^*$ for $\phi$ within the $\epsilon$-neighbourhood of $\vec{x}_0$ with Projected Gradient Descent (PGD)~\cite{madryDeepLearningModels2018} and using it in training. This approach is similar to adversarial training. However the perturbation found using PGD is only used as a counterexample for the logical constraint, and not for calculating the mean squared error loss term. Thus in turn adversarial examples do not improve the performance of the network; they are only used to make the network satisfy the constraint more. The code we use to achieve this in training is based on~\cite{flinkowComparingDifferentiableLogics2024}.

\section{Results and Observations}
We performed two variations of training of the suggested neural network on our collected data.
Firstly, we trained the network in a standard manner (called \emph{vanilla} in the following) using only our collected data.
Secondly, we trained our network on the collected data but added constraints (called \emph{constrained-training} in the following).
Both training runs were executed for 100 \emph{epochs} with a \emph{batch size} of 16.
\subsection{Performance}

\begin{figure}
  \centering
  \begin{subfigure}{.5\textwidth}
    \centering
    \includegraphics[width=.9\linewidth]{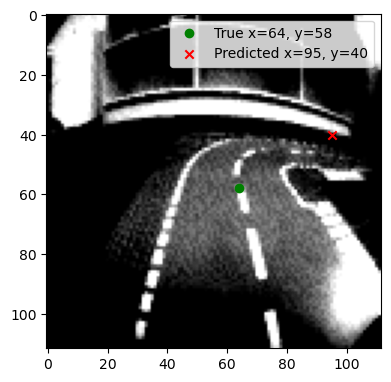}
    \caption{Vanilla}
    \label{fig:sub1}
  \end{subfigure}%
  \begin{subfigure}{.5\textwidth}
    \centering
    \includegraphics[width=.9\linewidth]{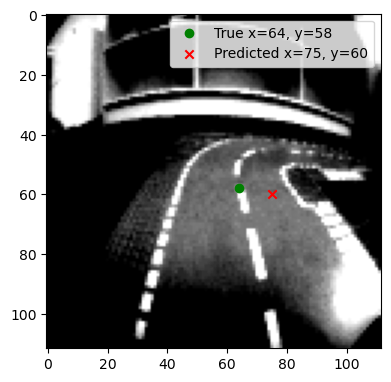}
    \caption{Constrained-training}
    \label{fig:sub2}
  \end{subfigure}
  \caption{Comparison of predictions of models trained without (vanilla) and with logical constraints (constrained-training) on an adversarial example in epoch 45 of 100.}
  \label{fig:performance_on_adv}
\end{figure}

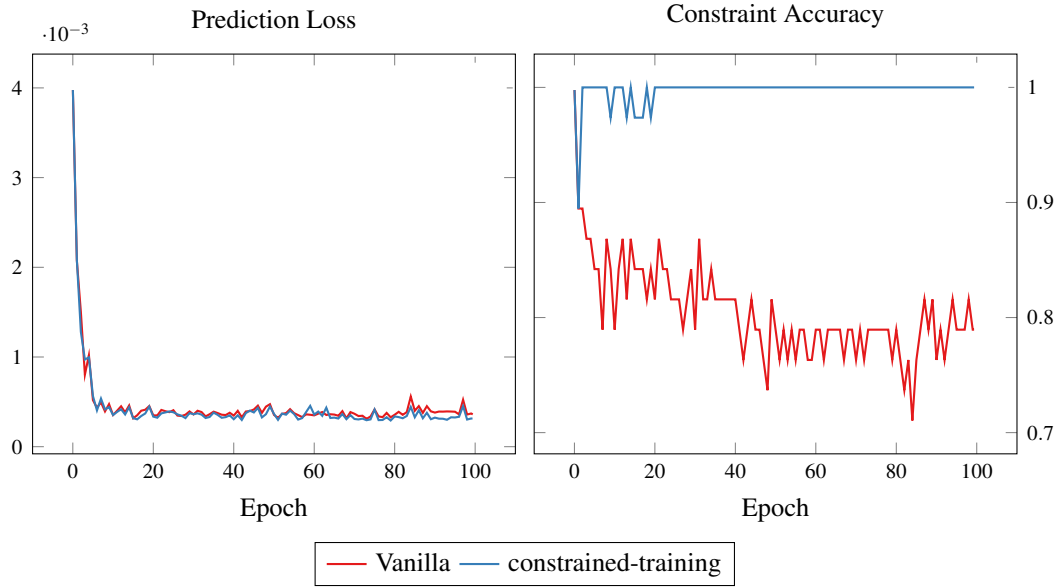
\begin{figure}
    \centering
    \begin{tikzpicture}[font=\footnotesize]
      \begin{groupplot}[group/results]
        \nextgroupplot[title={Prediction Loss}, yticklabel pos=left]
        \addplot table [y=Test-P-Loss] {Baseline.csv};
        \addplot table [y=Test-P-Loss] {GD.csv};
        
        \coordinate (c1) at (rel axis cs:0,1);
            \nextgroupplot[title={Constraint Accuracy},
              yticklabel pos=right,
              legend to name=full-legend
            ]
        \addplot table [y=Test-C-Acc] {Baseline.csv};
        \addplot table [y=Test-C-Acc] {GD.csv};
        \addlegendentry {Vanilla};
        \addlegendentry {constrained-training};    
        \coordinate (c2) at (rel axis cs:1,1);
      \end{groupplot}
      \coordinate (c3) at ($(c1)!.5!(c2)$);
      \node[below] at (c3 |- current bounding box.south) {\pgfplotslegendfromname{full-legend}};
    \end{tikzpicture}
    \caption{The prediction loss (i.e., mean squared error; lower is better) and constraint accuracy (higher is better) for a model trained only on data (vanilla) and a model trained on both data and logical constraints with differentiable logics (constrained-training).}
    \label{fig:accuracy_and_loss}
\end{figure}

To evaluate the performance, we look at prediction loss (i.e. mean squared error) and constraint accuracy (i.e. the number of times the constraint was satisfied out of all the predictions). \cref{fig:accuracy_and_loss,fig:performance_on_adv} illustrate that including constraints in the training process leads to improved performance on adversarial examples, as well as drastically improving constraint accuracy. However, as mentioned in \cite{giunchigliaDeepLearningLogical2022}, training with constraints does not guarantee their satisfaction. Instead, we aimed to obtain formal guarantees that both networks verify the constraints after training.

\subsection{Verification of Robustness}
We used $\alpha,\beta$-CROWN to verify that the network satisfies the robustness property in~\cref{eq:robustness}. However, we found that the tool was not suitable for the verification of regression tasks (as opposed to classification tasks) without modification. We also used DNNV~\cite{shriverDNNVFrameworkDeep2021}, a framework providing a unified interface to interact with verifiers such as Reluplex~\cite{katzReluplexEfficientSMT2017a}, Marabou~\cite{katzMarabouFrameworkVerification2019}, BaB~\cite{bunelUnifiedViewPiecewise2018}, ERAN~\cite{singhFastEffectiveRobustness2018,singhBoostingRobustnessCertification2018,singhAbstractDomainCertifying2019,singhSingleNeuronConvex2019}, and nnenum~\cite{bakNnenumVerificationReLU2021}.
DNNV provides a Python-based domain-specific language for expressing properties.
Our translation of the constraint \cref{eq:robustness} is shown in \cref{lis:dnnv}.
DNNV was not straightforward to install, requiring older versions of Python along with outdated requirements for packages such as numpy.
The only verifiers that we were able to install successfully within DNNV were BAB, ERAN, Reluplex and nnenum.
Verifying the property shown in \cref{lis:dnnv} led to an inconclusive result for ERAN, whereas Reluplex and BAB were not able to attempt verification due to unsupported operations, and nnenum returned a generic error.
The tool's output for these scenarios is shown in \cref{output:dnnv}.

\begin{lstlisting}[float=tb,language=Python,caption={The robustness property from \cref{eq:robustness} specified in DNNV's property specification language.},label=lis:dnnv]
from dnnv.properties import *

N = Network("N")
x = Image("data/image_0.npy")

epsilon = Parameter("epsilon", float, default=(48. / 255))
delta = Parameter("delta", float, default=0.1)

Forall(
    x_,
    Implies(
        ((x - epsilon) <= denormalise(x_) <= (x + epsilon)),
        (abs(N(x_)[0][0] - N(x)[0][0]) <= delta) &
        (abs(N(x_)[0][1] - N(x)[0][1]) <= delta)
    ),
)
\end{lstlisting}

\begin{lstlisting}[float=tb,caption={Output of DNNV for  verification of robustness property.},label=output:dnnv]
dnnv.verifiers.bab
  result: BabTranslatorError(Unsupported computation graph detected)
  time: 0.3495

dnnv.verifiers.eran
  result: unknown
  time: 4.9732

dnnv.verifiers.nnenum
  result: NnenumError(Return code: 1)
  time: 0.7476

dnnv.verifiers.reluplex
  result: ReluplexTranslatorError(Unsupported computation graph detected)
  time: 0.3622
\end{lstlisting}

\begin{figure}
  \centering
\includegraphics[width=.25\linewidth]{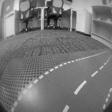}
    \caption{An image (converted to grey scale) from the LEGO dataset taken as a running example.}
    \label{fig:running_example}
\end{figure}

We also used the Matlab Toolbox for Neural Network Verification (NNV) \cite{tranNNVNeuralNetwork2020} for verification of the robustness property.
NNV provides a relatively simplier interface that accepts several neural network formats.
The tool has an \texttt {estimateNetworkOutputBounds} function that estimates lower and upper output bounds of the network, supporting a change in input within the specified bounds. For a small perturbation $\epsilon$ in input $X$, the input bounds are $X - \epsilon$ and $X + \epsilon$.  
For the input image in~\cref{fig:running_example}, \cref{table:t1} shows the lower and upper bounds ($x$, $y$ coordinates) estimated by NNV against the perturbations in the normalised input for different network training configurations. It can be observed from \cref{table:t1} that the \texttt {estimateNetworkOutputBounds} function does return any tighter bounds for the network's output, showing the sensitivity of the network to input changes.

\begin{table}
\centering
\caption{Lower and upper output bounds of the proposed neural network computed by NNV against perturbations in the normalised input image from Figure \ref{fig:running_example}.}
\begin{tabular}{|c|clll|}
\hline
\multirow{3}{*}{\textbf{\begin{tabular}[c]{@{}c@{}}Input Perturbations $\epsilon$\\ (Normalised Image)\end{tabular}}} & \multicolumn{4}{c|}{\textbf{Network Configuration}}                                              \\ \cline{2-5} 
                                       & \multicolumn{2}{c|}{Vanilla}                    & \multicolumn{2}{c|}{Constrained-training}                         \\ \cline{2-5} 
                                       & \multicolumn{1}{c|}{$x$} & \multicolumn{1}{c|}{$y$} & \multicolumn{1}{c|}{$x$} & \multicolumn{1}{c|}{$y$} \\ \hline
0.001 & \multicolumn{1}{l|}{{[}9 -- 107{]}} & \multicolumn{1}{l|}{{[}19 -- 97{]}} & \multicolumn{1}{l|}{{[}8 -- 107{]}} & {[}15 -- 97{]} \\ \hline
0.01  & \multicolumn{1}{l|}{{[}1 -- 112{]}}  & \multicolumn{1}{l|}{{[}1 -- 112{]}} & \multicolumn{1}{l|}{{[}1 -- 112{]}}  & {[}1 -- 112{]} \\ \hline
\end{tabular}
\label{table:t1}
\end{table}

\subsection{Lessons Learned}
Supporting formal guarantees for neural networks appears qualitatively different from traditional formal verification.
While formal verification techniques for neural networks are general enough to allow the verification of various properties over certain regions of the input space, in practice, finding these regions can be problematic---for example, requiring a network to be robust usually applies only for specific images, not all possible images.
As this is difficult to specify for all but the most simple, low-dimensional problems (such as ACAS Xu~\cite{katzReluplexEfficientSMT2017a}), verification is usually limited to verifying local properties at input points contained in the available data. This has important implications for the use of the network in formally verified systems---such as the lack of real guarantees for unseen data.

The architecture of the neural network has extensive consequences for the whole pipeline, from affecting the training time with differentiable logics, to making verification difficult or potentially impossible.
As explained in~\cite{tranVerificationDeepConvolutional2020}, max-pooling layers are typically too complicated for verification. The same applies to $\tanh$ layers, as many tools support only ReLU activation functions, fully connected, and convolutional layers, which explains why Reluplex and BaB failed to verify our property, as seen in~\cref{output:dnnv}.

 Regarding verification tools, the effort in installation, interfacing tools with the network to be verified, and getting the tool running in terms of computational resources is substantial. Tools were difficult to install due to their (sometimes outdated) dependencies, and the data and network often required conversion to an acceptable format to be compatible with the tools expectations. We note also that neural network verification tools focus primarily on networks performing classification rather than regression, as evidenced by the large number of examples provided for the earlier case.

Lastly, considering the autonomous navigation example in our case study and the dataset design, we would like to explore using further labelled data, e.g., the left and right edges of the track could be included in addition to the centre of the track. This opens up more interesting logical constraints to use in training.

\section{Conclusion}
In this paper  we presented our experience of creating a formally verified neural network for autonomous navigation. While we gained many insights into the currently available tools and approaches, the challenge of creating verification-friendly neural networks (in general and) for autonomous driving systems is still an open problem, requiring expertise from both the formal verification and the neural network communities.

We investigated the design and training of a neural network on a custom dataset for vision-based autonomous navigation in a fashion that integrates standard training methods and logic-based formal methods. In particular, we used differentiable logics to constrain training to yield networks satisfying
safety properties like robustness. This approach can be integrated into a pipeline in which the networks thus trained are then checked for compliance using formal verification of the post-training neural network. 


\bibliographystyle{eptcs}
\bibliography{references}

\end{document}

%% file: tikz_settings.tex
\usepackage{tikz}
\usepackage{pgfplots}
\usepackage{pgfplotstable}
\usepgfplotslibrary{colormaps, colorbrewer, groupplots}

\usepackage{amsmath}

\pgfplotsset{
  compat=1.18,
  cycle list/Set1,
  results/.style={
    width=8cm,
    xtick={0,20,...,100},
    tick label style={font=\scriptsize},
    legend cell align={left},
    legend style={
      at={($(0,0)+(1cm,1cm)$)},
      legend columns=3,
      anchor=center,
      align=center,
      font=\scriptsize
    },
    legend image post style={mark indices={}},
    enlargelimits=true,
    xlabel=Epoch,
    every axis plot post/.append style={
      thick,
      table/col sep=comma,
      table/x=Epoch,
    },
  },
  group/results/.style={
    group style={
      group size=2 by 1, 
      ylabels at=edge left,
      x descriptions at=edge bottom,
      horizontal sep=0.25cm,
    }, 
    results,
  },
}